\documentclass{iaus}
\usepackage{graphicx}

\title[Microlensing search for extrasolar planets] 
{Microlensing search for extrasolar planets: observational strategy, discoveries and implications}

\author[Arnaud Cassan, Takahiro Sumi, Daniel Kubas]   
{
Arnaud Cassan$^{1, 4}$\thanks{Based on the talk ``Microlensing search for extrasolar planets''.},
Takahiro Sumi$^{2, 5}$\thanks{Based on the talk ``Exoplanet Candidates from the MOA-II Microlensing Survey in 2007''.}, 
Daniel Kubas$^{3, 4}$ 
}

\affiliation
{
$^1$ ARI/ZAH Heidelberg University, Germany \\[\affilskip]
$^2$ Nagoya University, Japan \\[\affilskip]
$^3$ ESO, Chile \\[\affilskip]
$^4$ PLANET/RoboNET Collaborations \\[\affilskip]
$^5$ MOA Collaboration
}

\pubyear{2008}
\volume{xxx}  
\pagerange{119--126}
\jname{Title of your IAU Symposium}
\editors{A.C. Editor, B.D. Editor \& C.E. Editor, eds.}
\begin{document}

\maketitle

\begin{abstract}
  Microlensing has proven to be a valuable tool to search for extrasolar planets
  of Jovian- to Super-Earth-mass planets at orbits of a few AU. Since planetary signals
  are of very short duration, an intense and continuous monitoring is required. This
  is achieved by ground-based networks of telescopes (PLANET/RoboNET, $\mu$FUN)
  following up targets, which are identified as microlensing events by single dedicated telescopes (OGLE, MOA).
  Microlensing has led to four already published detections of extrasolar planets, 
  one of them being OGLE~2005-BLG-390Lb, a
  planet of only $\sim5.5\,M_\oplus$ orbiting its M-dwarf host star at $\sim2.6$ AU. 
  Very recent observations (May--September 2007) provided four more planetary candidates,
  still under study, that will double the number of detections.
  For non-planetary microlensing events observed from 1995 to 2006 we compute
  detection efficiency diagrams,
  which can then be used to derive an estimate of the Galactic abundance of cool planets
  in the mass regime from Jupiters to Sub-Neptunes.
\keywords{Extrasolar planets, Gravitational microlensing}
\end{abstract}

\firstsection 

\section{Introduction}

A Galactic microlensing event occurs when a massive compact intervening object (the lens)
deflects the light coming from a more distant background star (the source). 
It leads to an apparent flux brightening (or magnification)
of the source star. In a typical
scenario, the source belongs to the Galactic Bulge, while the lens can be part either of the Bulge (2/3 of the events) or the
Disk (1/3 of the events) population. Mao \& Paczy\'nski (1991) were the first to suggest that microlensing 
could provide a powerful tool to search for extrasolar planets at distances of a few kpc, provided
a continuous monitoring of Bulge stars. 
Since the detection of planets by microlensing does not rely on their light but
on their mass, the planetary host mass function basically follows the stellar
mass function of the Galaxy, implying that planet hosts are preferably low-mass K to M dwarfs.

The relative motion between source, lens and observer ($\sim15\,\mu$as/day) induces a variation of the magnification factor
with time, with a typical time scale of 
$t_{\rm E} \simeq 40 \times \sqrt{M_*/M_\odot}$ days, assuming 
a source and lens distance of respectively $8.5$ and $6.5$~kpc.
The duration of the planetary light curve signal then scales as
$t_{\rm p} \approx 2\sqrt{q}\times t_{\rm E}$, where $q$ is the planet-to-star mass ratio, which means
few days for a giant planet to only few hours for Neptune- to Earth-mass planets. 

The microlensing method is remarkable in the sense that it probes a domain in the planet mass-orbit diagram that is mainly out of reach of other techniques, for it is mostly sensitive to Jovian- down to Earth-mass planets (e.g. Bennett \& Rhie 2002)
with orbits of $\sim 1-10$~AU, at several kpc.

\section{Alerts and follow-up of microlensing events}

Microlensing surveys (OGLE \& MOA) currently monitor more than $\sim 10^8$ 
Galactic Bulge stars by 1-2 m class telescopes on a daily basis to find and
alert microlensing events.
The second phase of MOA, MOA-II, carries out survey observations
toward the Galactic Bulge to find exoplanets via microlensing using 
a 1.8m telescope in New Zealand. We observe our target fields 
($\sim 50 ~$deg$^2$) very frequently (10 to 50 times/night) and analyze data in
Real-time to issue microlensing alerts. This high cadence is specifically
designed to find the short duration time signatures characteristic of
planets orbiting the lens star.  In 2007 realtime
event monitoring started in order to search for planetary signatures in ongoing microlensing
events. Each new data point on the light curves is available within 5
minutes  after image exposure. During 2007, MOA has detected around 500
microlensing events. 

While single alert telescopes are able to identify and follow microlensing planetary
candidates, they somehow suffer from gaps in the data coverage. Network of telescopes,
as operated by the PLANET/RoboNET or $\mu$FUN collaborations perform 
a ``round-the-clock'' monitoring of a reduced number of selected targets, which
significantly increases the planet detection efficiency.
For example, with currently five 1m-class telescopes located in Chile, two in South Africa, Australia and Tasmania,
as well as using three robotic telescopes,
PLANET/RoboNET currently has unequaled capability
for covering microlensing events, by minimizing
data gaps in which planetary signatures could hide.

Recent efficient and interactive communication between alert and network collaborations 
has played a major role in improving the ability to quickly focus on suspected planetary
signal in many events.

\section{Results}

\subsection{Detections and new planetary candidates}

As of now four extrasolar planets detected by microlensing are reported in the literature.
Among them two giants of few Jupiter masses,
MOA~2003-BLG-53Lb  (\cite{mb53Lb}) and OGLE~2005-BLG-071Lb (\cite{ob071Lb}),
as well as two Super-Earth-mass planets, OGLE~2005-BLG-169Lb (\cite{ob169Lb})
and OGLE~2005-BLG-390Lb (\cite{ob390Lb}), the latter being one the lightest ever discovered 
planets, with only $\sim5.5\,M_\oplus$ and a wide orbit of $\sim2.6$ AU.
This first detection of a cool rocky/icy sub-Neptune mass planet
has thus opened a new observing window for the exoplanet field.

Very recent observations (May--September 2007) have revealed four more 
planetary candidates which are in the process of final analysis and are expected to be
published in 2008. These new detections will double the number of detections and should
provide a better understanding of the statistical properties of the microlensing planet population and
also help to optimize  he detection efficiency of the current observational set up (see \ref{sec:effdet}).

\subsection{Limits on the multiplicity of planetary systems}

\begin{figure}[!ht]
  \begin{center}
    \includegraphics[width=10cm]{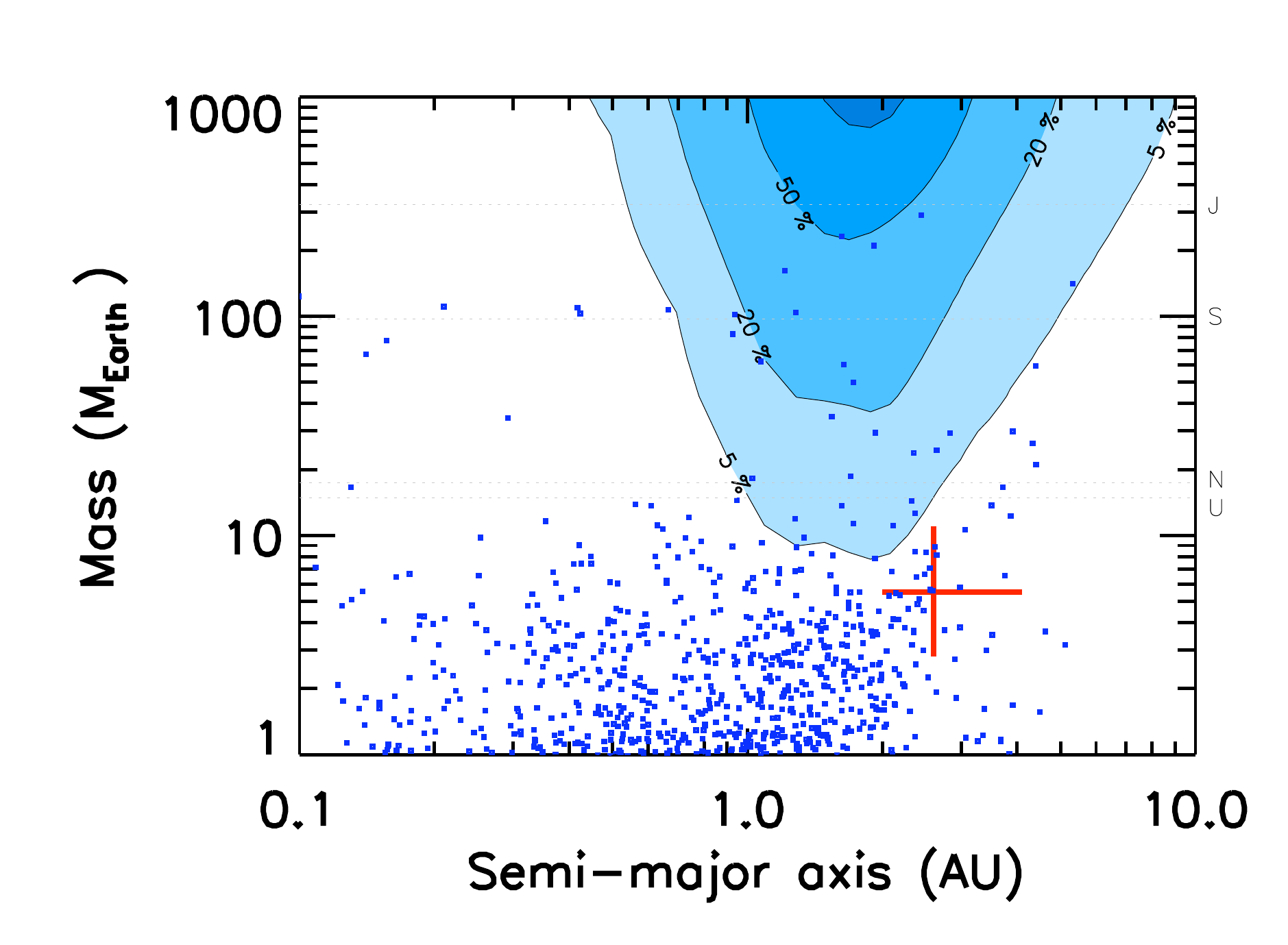}
    \caption{
   Detection efficiency for an additional planet orbiting
OGLE-2005-BLG-390L as a function of its orbital separation and
mass.  Efficiency contours of $5\%$, $20\%$, $50\%$ and $70\%$ are shown. 
The cross marks the median values for the properties of OGLE 2005-BLG-
390Lb along with $1~\sigma$ confidence intervals and the dashed horizontal
lines mark the masses of Jupiter (J), Saturn (S), Neptune (N) and
Uranus (U) for comparison. The blue dots represent the predicted final
distribution of a seed of $2\times10^{4}$ planetary cores around an M-dwarf
of $0.2~M_{\odot}$ resulting from a core-accretion model assuming inefficient
migration (taken from Fig. 9b of Ida \& Lin (2005)). }
   \label{fig:ob390furth}
  \end{center}
\end{figure}

Microlensing also allows the direct detection of a multi-planetary system (see the contribution 
from D.~Bennett, this volume), as well as of circum-binary planets.

When only a single planet is detected, it is still possible to put limits on the presence of further planets in the 
microlensing event, such as in the case of OGLE 2005-BLG-390 (\cite{further390p}), presented
in Fig.~\ref{fig:ob390furth}.
Although the detection efficiency depends strongly on the reached peak magnification, a good data coverage
can result in significant detection sensitivities even for  low peak magnification events.  In  this particular case
one finds  that more than $50\%$ of potential planets with a mass in excess of 
$1~M_J$ between $1.1$ and $2.3$ AU around  OGLE 2005-BLG-390L would have revealed their existence, which was however not observed. For gas giant planets above $3~M_J$ in orbits between $1.5$ and
$2.2$ AU, the detection efficiency exceeds $70\%$. 
Furthermore we find a detection probability for an OGLE-2005-BLG-390Lb-like planet, given an idealization of the
microlensing technique, to be around $1-3~\%$. 
In agreement with current planet formation theories this quantitatively supports the prediction
that sub-Neptune mass planets are common around low mass stars.

\subsection{Detection efficiencies} \label{sec:effdet}

Apart from the detection of planets a main goal of microlensing observations is to
estimate the planet detection efficiency in order to  put constraints  the Galactic planet abundance.
From 42 densely monitored events between 1997 and
1999, PLANET was able to provide the first significant upper abundancy limit of Jupiter- 
and Saturn-mass planets around M-dwarfs, namely that less than $1/3$ of the lens
stars have Jupiter-mass companions at orbital radii between $1.5$ and
$4$~AU, and less than $2/3$ have Saturn-mass companions for the same
range of orbital radii, assuming circular orbits (\cite{effgau}).

\begin{figure}[!ht]
  \begin{center}
    \includegraphics[width=10cm]{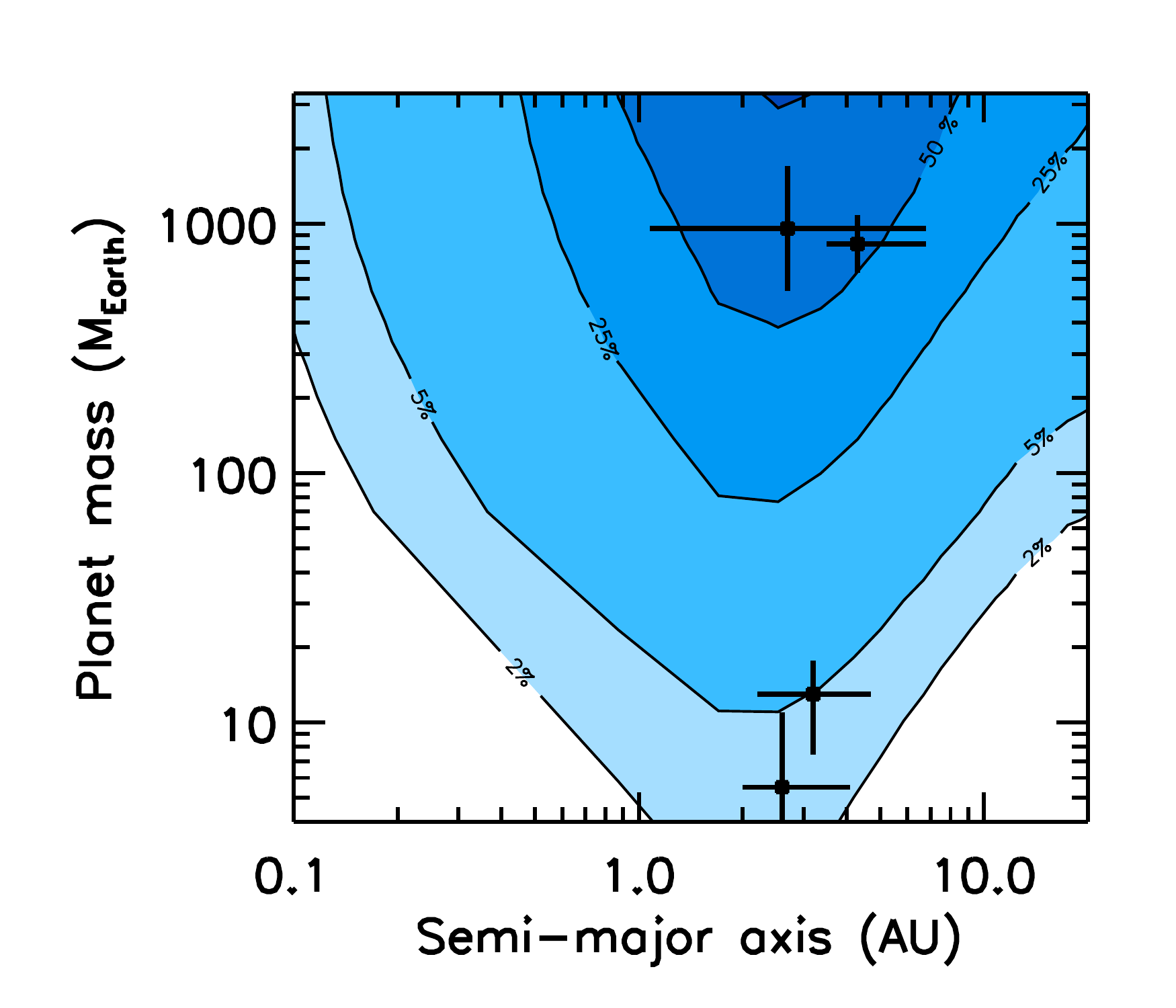}
    \caption{PLANET detection efficiency from the 2004 season (preliminary diagram), as a function
    of planet mass and orbital separation. The crosses are the detected planets with their parameter
    error bars.}
    \label{fig:eff}
  \end{center}
\end{figure}

By using an adequate Galactic model for the distribution of lens masses and velocities (\cite{GalMod}),
we aim to pursue and improve the study, moreover taking into account 
more than ten years of observations (\cite{moai}). 
The Fig.~\ref{fig:eff} shows a preliminary planet detection efficiency diagram, computed
from well-covered events of the 2004 season.

\section{Summary and prospects}

Microlensing has proven to be a robust method to search for extrasolar planets at large separations from their
parent stars ($\sim 1-10$~AU). It is sensitive to masses down to the mass of the Earth using ground based
telescopes and even capable to  detect planets of a  few fractions of Earth masses when considering space-based 
telescope scenarios.    

Microlensing is also very well-suited for statistical studies on  planet abundance in the Galaxy.
In fact, the method is by essence not limited to our close solar neighborhood or to a particular
type of host stars.



\end{document}